\documentclass[prl,aps,showpacs,superscriptaddress,amsmath,amssymb,floatfix,twocolumn]{revtex4-1}
\usepackage{graphicx}
\usepackage{color}
\usepackage{young}
\usepackage{youngtab}
\usepackage[colorlinks,bookmarks=false,citecolor=blue,linkcolor=red,urlcolor=blue]{hyperref}
\usepackage{amsfonts}
\usepackage{amsmath}
\usepackage{times}
\usepackage{amssymb}
\usepackage{changes}

\begin{document}
\title{Chiral spin liquids in triangular lattice $SU(N)$ fermionic Mott insulators with artificial gauge fields}
\date{\today} 

\author{Pierre Nataf}
\affiliation{Institute of Theoretical Physics, Ecole Polytechnique F\'ed\'erale de Lausanne (EPFL), CH-1015 Lausanne, Switzerland}
\author{Mikl\'os Lajk\'o}
\affiliation{Institute for Solid State Physics, University of Tokyo, Kashiwa 277-8581, Japan}
\author{Alexander Wietek}
\affiliation{Institut f\"ur Theoretische Physik, Universit\"at Innsbruck, A-6020 Innsbruck, Austria}
\author{Karlo Penc}
\affiliation{Institute for Solid State Physics and Optics, Wigner Research
Centre for Physics, Hungarian Academy of Sciences, H-1525 Budapest, P.O.B. 49, Hungary}
\affiliation{Department of Physics, Budapest University of Technology and Economics, 1111 Budapest, Hungary}
\author{Fr\'ed\'eric Mila}
\affiliation{Institute of Theoretical Physics, Ecole Polytechnique F\'ed\'erale de Lausanne (EPFL), CH-1015 Lausanne, Switzerland}
\author{Andreas M. L\"auchli}
\affiliation{Institut f\"ur Theoretische Physik, Universit\"at Innsbruck, A-6020 Innsbruck, Austria}

\begin{abstract}
We show that, in the presence of a $\pi/2$ artificial gauge field per plaquette, Mott insulating phases of ultra-cold fermions with $SU(N)$ symmetry and one particle per site generically possess an extended chiral phase with intrinsic topological order characterized by a multiplet of $N$ low-lying singlet excitations for periodic boundary conditions, and by chiral edge states described by the $SU(N)_1$ Wess-Zumino-Novikov-Witten conformal field theory for open boundary conditions. This has been achieved by extensive exact diagonalizations for $N$ between $3$ and $9$, and by a parton construction based on a set of $N$ Gutzwiller projected fermionic wave-functions with flux $\pi/N$ per triangular plaquette. Experimental implications are briefly discussed.
\end{abstract}


\maketitle

The search for unconventional quantum states of matter in realistic models of strongly correlated systems
has been an extremely active field of research over the last 25 years. Mott insulating phases in which 
charge degrees of freedom are gapped have been argued to potentially host several families of quantum
spin liquids ranging from Resonating Valence Bond $\mathbb{Z}_2$ quantum spin liquids~\cite{RokhsarKivelson1988,MoessnerSondhi2001,Ralko2005} to $U(1)$ algebraic spin liquids~\cite{AffleckMarston1988,Ran2007,Corboz_honeycomb_2011}
and chiral spin liquids~\cite{Kalmeyer1987,Wen1989a,Wen1989b,Greiter2007,Messio2012,He2014,Gong2014,Bauer2014}. The topological properties of these phases have attracted a lot of attention due to 
their potential impact on the implementation of quantum computers\cite{NayakRevModPhys.80.1083}. 

Cold atoms open new perspectives in that respect. In particular, alkaline rare earths allow
to realize $SU(N)$ Mott phases with $N$ as large as 10~\cite{gorshkov2010,Scazza2014,Pagano2014,Zhang2014}, and if a chiral phase can be stabilized, its low-energy 
theory is expected to be the $SU(N)$ level $k=1$ Chern-Simons theory. 
The first proposal of a chiral phase in this context goes back to the work of Hermele et al \cite{HermelePRL2009,HermeleGurarie2011}, 
who showed that a mean-field approach leads to the stabilization of chiral
phases on the square lattice in the limit of large $N$ and large number of particles per site $m$ with 
$N/m$ integer and $\geq 5$. The same mean-field applied to $SU(6)$ on the honeycomb lattice with
one particle per site has also led to the prediction of a chiral state,  with a competing plaquette
state very close in energy \cite{SzirmaiPRA2011}. More recently Ref.~\cite{Chen2015} suggested the stabilization
of $SU(N)$ chiral spin liquids on the square lattice using static synthetic gauge fields, based on a slave-rotor
mean-field approach. In all theses cases, the results call for further investigation with methods that go
beyond mean-field theory. 

\begin{figure*}[t]
\begin{center}
\includegraphics[width=\linewidth]{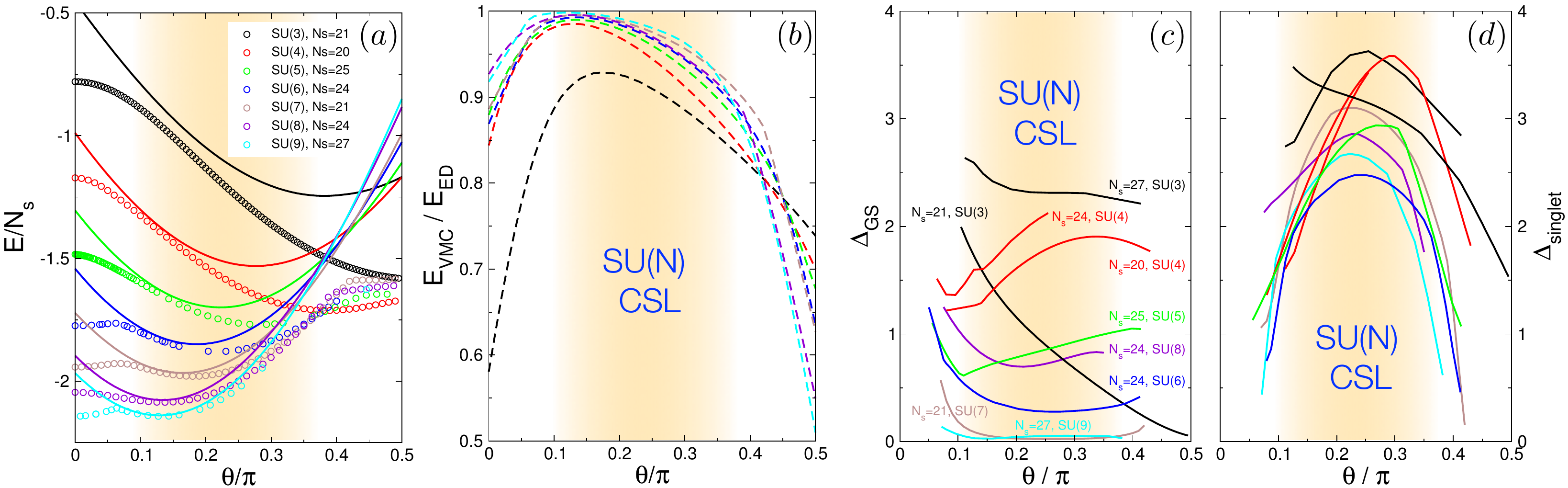}
\caption{Panel (a): Ground state energy per site as a function of $\theta$ for various $N$ and $N_s$. Open symbols (full lines) denote ED (VMC) results.
(b): Quality of the VMC wave function as measured by the ratio $E_\mathrm{VMC}/E_\mathrm{ED}$. (c): Energy splitting among the expected $N$ singlet states
forming the ground space manifold of a $SU(N)$ chiral spin liquid. (d) Energy gap from the ground state to the first excited singlet state which is not part of the expected
ground space manifold.}
\label{fig:energy_spectra}
\end{center}
\end{figure*}

In this Letter, we show that the ground state of the Mott phase of $N$-color fermions on the triangular
lattice with one particle per site is a $SU(N)$ chiral spin liquid in a large parameter range if the system is subject to a
static artificial gauge field with flux $\pi/2$ per triangular plaquette. The starting point is the $SU(N)$ Hubbard Hamiltonian 
\begin{equation}
H=-t\sum_{\langle i,j \rangle}\sum_{\alpha=1}^{N} (e^{\phi_{ij}}
c^\dagger_{i,\alpha}c^{\phantom \dagger}_{j,\alpha}+H.c.)
+U \sum_{i,\alpha<\beta} n_{i,\alpha} n_{i_\beta}
\label{eqn:Hubbard}
\end{equation}
where the phases $\phi_{ij}$ are chosen in a such a way that the (gauge-invariant) flux through each
triangular plaquette is equal to $\pi/2$. Then, at a filling of one particle per site, and for large enough $U/t$, the effective model
is an $SU(N)$ Heisenberg model with local spins in the fundamental representation of $SU(N)$ endowed with real pairwise 
permutations and purely imaginary three-site permutations defined by the Hamiltonian \cite{MotrunichPRB2006, SenChitraPRB1995}
\begin{equation}
H =J\sum_{\langle i,j \rangle} P_{ij} +K_3 \sum_{(i,j,k)} (iP_{ijk}+ h.c.)
\label{eqn:Heisenberg}
\end{equation}
where the sum over $(i,j,k)$ runs over all triangular plaquettes, and $P_{ij}$ and $P_{ijk}$ are circular permutation operators. 
To second order, the amplitude of the pairwise permutation is simply given by $J=2t^2/U$, while
the 3-site permutation appears at third order in perturbation theory with  $K_3=6t^3/U^2$~\cite{PhysRevB.87.205131}. 
In the following, we will discuss the properties of the model \eqref{eqn:Heisenberg} as a function of $J$ and $K_3$ using the
parametrization $J=\cos \theta$ and $K_3=\sin \theta$. We will discuss the experimental prospects of
realizing this Hamiltonian towards the end of the manuscript. It is interesting to note that parent Hamiltonians for $SU(N)$ chiral 
spin liquids have been proposed recently~\cite{Tu2014328,Bondesan2014483}. They include both the two-site permutations and the
imaginary part of the cyclic three-site permutations, but also in addition the real part of the cyclic three-site permutations, which
we omit. The range of the terms in the parent Hamiltonians are however not restricted to nearest neighbor or the elementary
triangular plaquette only, but the amplitudes depend in a power-law fashion on the distances among the two or three spins. While
there are some structural similarities, it is not obvious that the spatially compact Hamiltonian~\eqref{eqn:Heisenberg} features CSL phases.  
It is the goal of this Letter to provide compelling numerical evidence, based on large-scale Exact Diagonalizations (ED) and Gutzwiller
projected parton wave functions, that the above Heisenberg Hamiltonian indeed features extended regions of $SU(N)$ CSLs for all
values of $N=3$ to $9$ considered here.

\paragraph{Exact diagonalizations --} 

We start by investigating finite periodic triangular lattice clusters as a function of $\theta$ for
various values of $N$.
We focus on the range $\theta \in [0,\pi/2]$ in the following. $\theta >\pi/2$ is likely to be dominated
by ferromagnetism, while $\theta<0$ yields the time-reversed, but otherwise identical physics as $-\theta$. 
For small values of $N=3,4$ we used the standard ED approach employing
all the space group symmetries, while only considering the individual color conservation, corresponding to an abelian subgroup of $SU(N)$.
For all other $N$ a recently developed ED approach by two of the authors~\cite{NatafMilaPRL2014}, exploiting the $SU(N)$ symmetry at the expense of spatial symmetries,
is currently the only way to address these systems within ED. Depending on $N$, the largest system sizes $N_s$ range from 21 to 27 lattice sites.

In Fig.~\ref{fig:energy_spectra}(a) we plot the ED results for the energy per site of the ground state as a function of $\theta$ for all considered $N$ (open symbols). While the
curves for $N\lesssim 5$ look rather smooth at first sight, it is visible that the energy per site displays kinks around $\theta/\pi\sim 0.05-0.1$ and at
$\theta/\pi\sim 0.35-0.4$ for $N=6$ to $9$. For comparison we plot the energy expectation value of parameter-free Gutzwiller projected 
chiral spin liquid model wave functions for all values of $N$ (full lines). We will discuss the properties of these wave functions in a moment.
Interestingly, these model wave functions have very competitive energies, especially in the $\theta$ region slightly above the first kink. 
For a quantitative comparison we show in Fig.~\ref{fig:energy_spectra}(b) the ratio of the variational energy divided by the ED ground state 
energy. It is impressive that for $N$ beyond 3 the best ratio exceeds $0.98$ for the system sizes considered. So the picture so far is that the
small and large $\theta$ regimes for all considered $N$ are most likely other phases, while the intermediate region could harbour chiral spin liquids. 

$SU(N)$ chiral spin liquids are intrinsically topologically ordered: They exhibit a non-trivial ground state degeneracy on the torus~\cite{HermeleGurarie2011}
and fractional excitations. The ground state degeneracy on the torus is expected to be $N$ for these particular states 
with $N$ different abelian anyons~\cite{HermelePRL2009,HermeleGurarie2011}. In our numerical simulations, we can detect this degeneracy by investigating
the low-energy spectrum on samples with a total number of lattice sites $N_s$ that is an integer multiple of $N$. In Fig.~\ref{fig:energy_spectra}(c) we display
the energy spread $\Delta_\mathrm{GS}$ of these $N$ expected ground states for different $N$ as a function of $\theta$. As a general trend we observe that the splitting reduces 
significantly as we increase $N$. On the other hand several samples still show a substantial splitting. Naively one would expect a simple exponential suppression
of the splitting with system size, however in the related context of fractional Chern insulators a more subtle dependence of the ground space splitting 
on the actual shape of the clusters has been observed and rationalized~\cite{AML_FCI_2013}. We think that similar considerations apply here as well.

Finally we also measure the gap $\Delta_\mathrm{singlet}$ from the absolute ground state to the first singlet level that is not part of the expected ground state 
manifold. This is a measure for the excitation gap in the gapped chiral spin liquid states. In Fig.~\ref{fig:energy_spectra}(d), one observes an approximate dome-shaped
behaviour of this gap for all $N$, and furthermore this gap seems to depend only weakly on $N$. The approximate region in $\theta$ where the $N$-fold 
ground state degeneracy splitting is small compared to the excitation gap (for large $N$) is indicated as a shaded region in all the panels, and indicates a rough 
stability region for the $SU(N)$ chiral spin liquids on the triangular lattice. One should note however that the precise extent of the chiral spin liquids for small $N$ is an open question at this point.

\paragraph{Variational~parton~approach --}
\begin{figure}
\centerline{\includegraphics[width=0.75\linewidth]{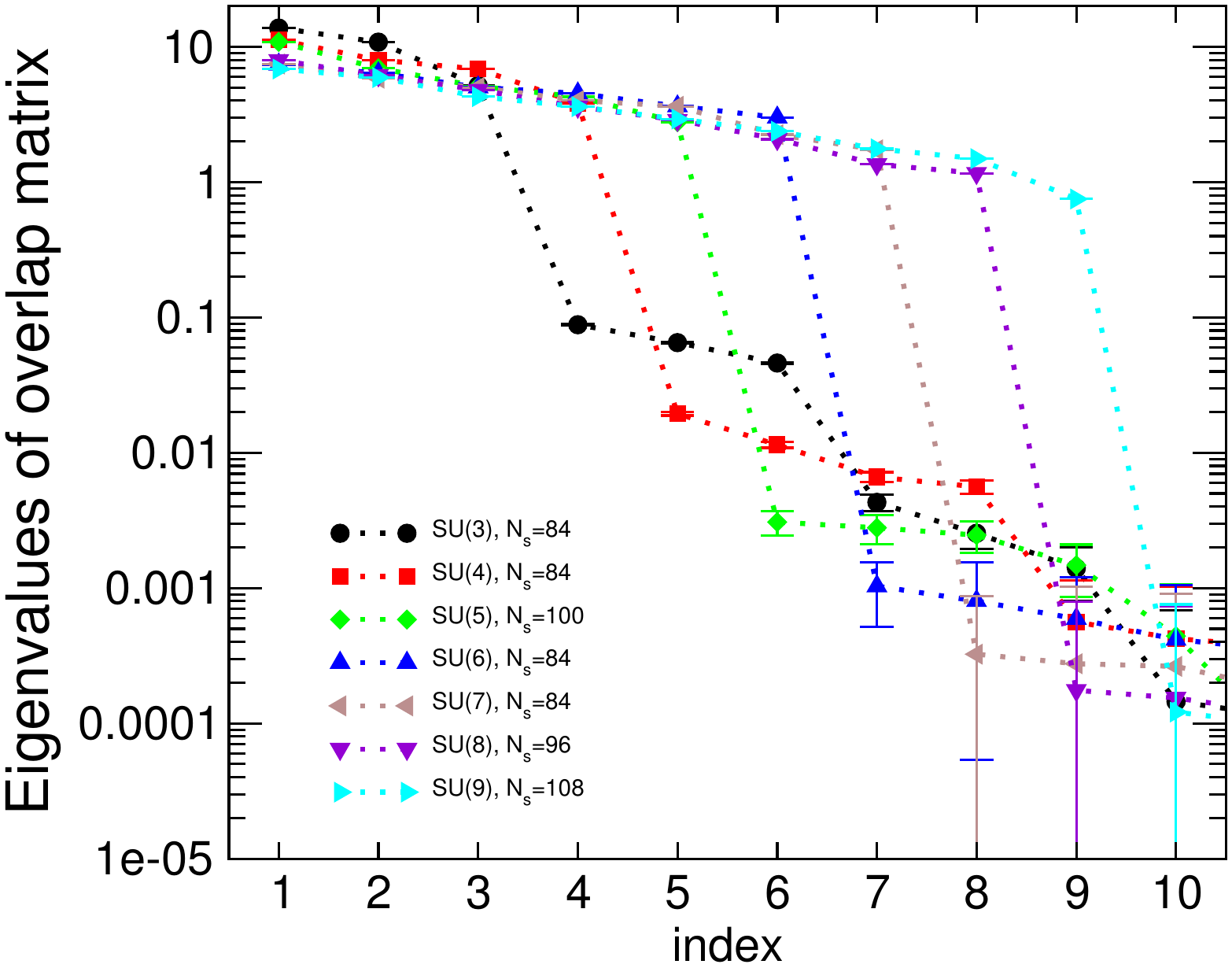}}
\caption{VMC ground space degeneracy: ordered sequence of eigenvalues of the overlap
matrix of Gutzwiller projected wave functions with 30 different values of threaded flux. 
The overlap matrix has precisely $N$ large eigenvalues for an $SU(N)$ chiral spin liquid.}
\label{fig:vmc_subspace}
\end{figure}

An appealing way to describe the $SU(N)$ chiral spin liquids is to use a parton-based mean field approach~\cite{Baskaran1987973,baskaran1988gauge,PhysRevB.38.745,PhysRevB.38.2926,PhysRevLett.76.503,senthil2000z,HermelePRL2009,HermeleGurarie2011}, complemented
with a Gutzwiller projection. The idea is to fractionalize the elementary spin degree
of freedom into fermionic spinons (partons) with $N$ flavors. For an exact description a dynamical gauge field needs to enforce the physical constraint
of one fermion per site. At the mean-field level however it is sufficient to specify the band structure and filling of the fermionic spinons.
In the $SU(N)$ chiral spin liquids of interest here, the spinon band structure consists of $N$ bands, where the lowest band is completely filled for all 
$N$ flavors and separated by a gap from the other bands. In addition this band is required to have Chern number $\pm1$. For the triangular lattice 
we use a Hofstadter-type tight-binding Hamiltonian with a uniform flux of $\pi/N$ per triangular plaquette~\footnote{Note that at this stage the flux per 
plaquette is unrelated to the flux per plaquette in the original Fermi-Hubbard Hamiltonian~\eqref{eqn:Hubbard}}, fulfilling the requirements on the band
structure. This mean-field state can now be turned into a valid spin wave function by the application of an exact Gutzwiller projection, enforcing the 
presence of exactly one fermionic spinon per site. Such a wave function can be handled by Variational Monte Carlo (VMC) techniques, and in particular
one can easily calculate the energy of the Hamiltonian~\eqref{eqn:Heisenberg} on rather large lattices. The VMC energies displayed in 
Fig.~\ref{fig:energy_spectra}(a),(b) have been obtained this way. 

\begin{figure}
\centerline{\includegraphics[width=0.75\linewidth]{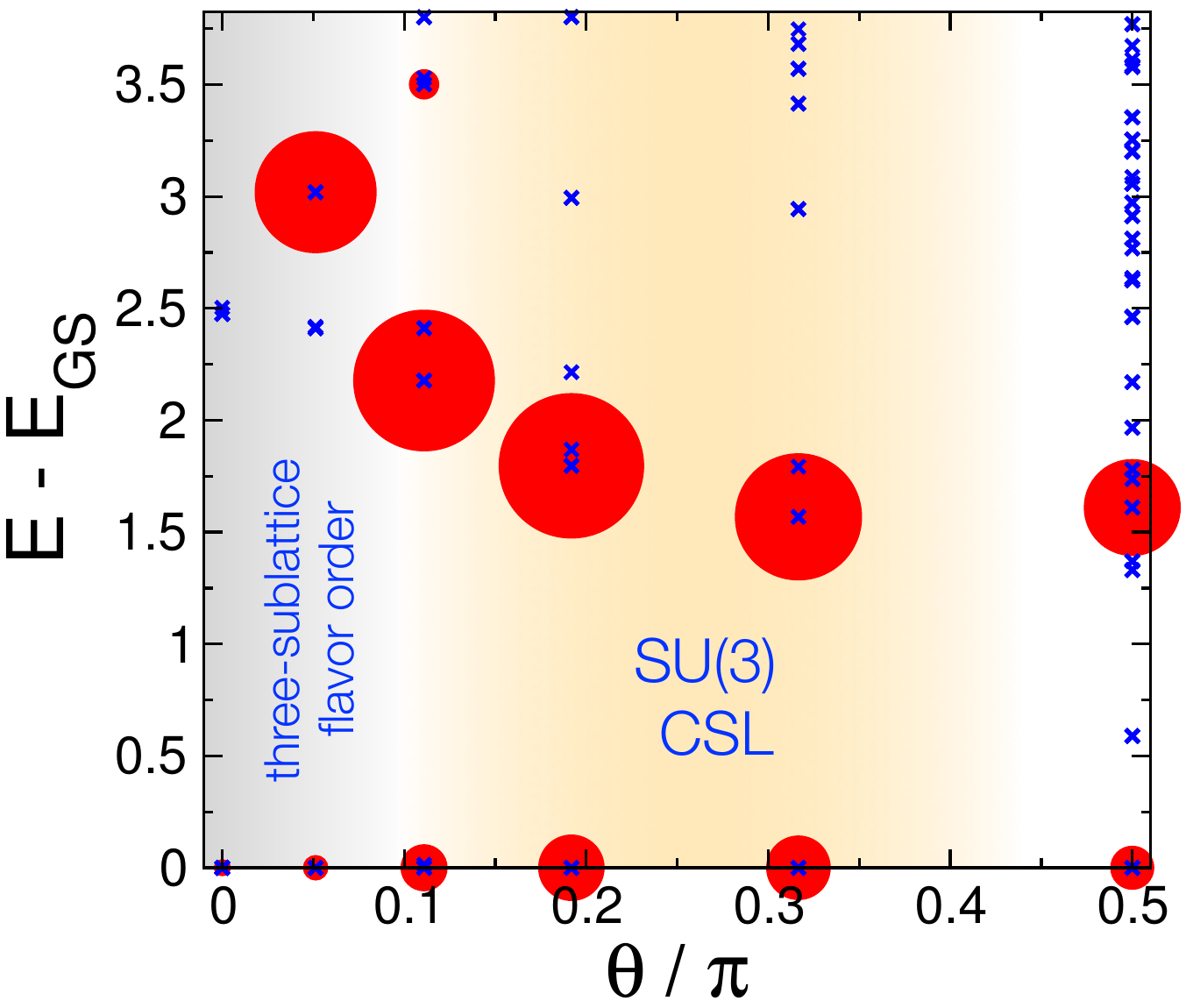}}
\caption{Summed squared overlaps of the VMC model wave functions with ED eigenstates for $N=3$ and $N_s=12$. The
blue crossed denote ED eigenstates, while the diameter of the filled red circles denotes the total squared overlap on those
eigenstates. In the best case the summed overlaps on the lowest three ED eigenstates (degeneracy $1+2$) account for over 90\% of the total weight.}
\label{fig:overlap_vmc_ed}
\end{figure}

The next question is how the VMC approach is able to account for the non-trivial ground
state degeneracy on the torus.  It turns out that by threading flux through the non-contractible loops around the torus, 
one is able to span an $N$-dimensional subspace of Gutzwiller projected wave-functions, with almost identical local properties on finite
lattices. From the viewpoint of topological order this corresponds to a charge pumping procedure, where one cycles through the $N$ different ground states
by threading different anyonic flux through the interior of the torus. These concepts have recently been explored in the context of $SU(2)$ CSL on several
lattices~\cite{PhysRevB.84.075128,Tu2013,WietekLaeuchli2015}. We have checked in Fig.~\ref{fig:vmc_subspace} that the subspace of wave functions 
spanned by using $30$ different boundary conditions at the mean-field level leads to a robust rank-$N$ overlap matrix, therefore corroborating the expectation
of an $N$-fold degenerate ground state manifold in the thermodynamic limit also at the VMC level.

Since the variational energies for $SU(3)$ turned out not to be very competitive, as shown in Fig.~\ref{fig:energy_spectra}(a)/(b),
we explicitly calculated the overlaps of individual ED eigenstates of the Hamiltonian~\eqref{eqn:Heisenberg} with the three orthogonal
Gutzwiller wave functions obtained on the same system size. In Fig.~\ref{fig:overlap_vmc_ed}  we plot the summed squared overlap of all three wave functions 
(diameter of filled circles) with the ED eigenstates (crosses) as a function of $\theta$. Here we consider a $N_s=12$ site system, where the
momenta of the three ED ground states in the chiral spin liquid phase are at the zone center (one) and at the corners of the Brillouin zone (twofold degenerate). Around $\theta=0$
the SU(3) triangular lattice Heisenberg model is in a three-sublattice flavor ordered state~\cite{Lauchli2006,Bauer2012}, however in the region around $\theta/\pi\sim 0.25$, the three lowest ED eigenstates indeed have sizeable overlap with the VMC model wave functions, thereby underlining the presence of an
$SU(3)$ chiral spin liquid for sufficiently large values of $\theta$ also for $N=3$.

\begin{figure*}[t]
\begin{center}
\includegraphics[width=\linewidth]{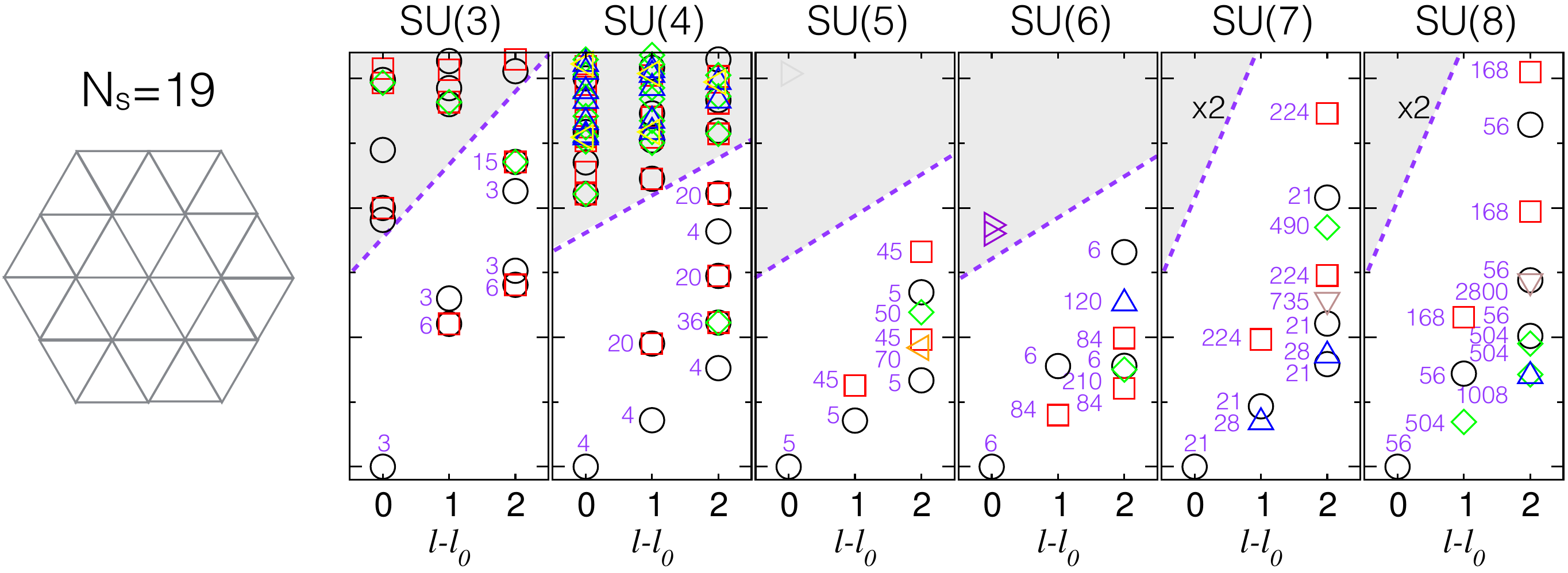}
\caption{Edge states in  $SU(N)$ chiral spin liquids: the leftmost panel displays the $N_s=19$ sites triangular cluster with
open boundary conditions used. In the various other panels we exhibit the low energy spectrum as a function of the angular momentum around
the central site ($\l_0$ denotes the ground state angular momentum). The chiral edge states are clearly visible, with a characteristic $SU(N)$
multiplet structure, which corresponds to a particular sector of a chiral $SU(N)_1$ Wess-Zumino-Novikov-Witten conformal field theory.
The analytical predictions are indicated by the dimensions of the $SU(N)$ multiplets and can be found in Tab.~\ref{tab:counting_sun_1_Ns_19} of
the supplementary material.}
\label{fig:edge_spectra}
\end{center}
\end{figure*}

\paragraph{Edge states --}
Another hallmark of chiral topological phases is the presence of chiral edge modes in the energy spectrum of systems with a boundary.
It has been understood that the characteristic energy level structure of the edge excitations as a function of the momentum along the 
boundary serves as a fingerprint of the type of topological order realised in the bulk~\cite{Wen2004}. The $SU(N)$
CSLs considered here are expected to exhibit a chiral edge energy spectrum described by the $SU(N)_1$ Wess-Zumino-Novikov-Witten (WZNW) 
conformal field theory (CFT)~\cite{HermeleGurarie2011}. This is the same CFT that governs the low-energy spectrum of well-studied 
one-dimensional critical $SU(N)$ spin chains~\cite{Sutherland1975,Affleck1986409,Tu2014328,Bondesan2014483}. 

In order to test this hypothesis numerically, one has to design a setup where one can detect the edge modes in a clean way. Starting from a torus
a natural way would be to cut the torus open into a cylinder. This geometry however has two independent, counter-propagating edges, making
a clean analysis difficult, given the system sizes $N_s$ accessible to ED. We therefore choose to emulate a disk geometry by considering the
specific $N_s=19$ site triangular lattice with open boundary conditions depicted in the left panel of Fig.~\ref{fig:edge_spectra}. Such a lattice might
actually be built in future ultracold atom experiments with optical lattices and a tight confining potential. This sample still has a sixfold rotation axis 
about the central site, yielding an angular momentum quantum number which we use to plot the energy spectrum. The energy spectrum of the disc 
has no topological ground state degeneracy,
but features gapless edge modes which typically propagate only in one direction. The precise multiplet structure of the edge modes depends on the
anyonic sector. In our setup this sector can be simply labeled as $a=(N_s\mod{N})$. In Tab.~\ref{tab:counting_sun_1_Ns_19} of the supplementary material
we have compiled 
the $SU(N)_1$ WZNW CFT predictions for the different irreducible representations of $SU(N)$ which appear at a given excitation energy, here qualitatively
labeled by the excess angular momentum $l-l_0$. In the remaining panels of Fig.~\ref{fig:edge_spectra} we display the actual ED energy spectrum of the
Hamiltonian~\eqref{eqn:Heisenberg} for a fixed value of $\theta/\pi=0.25$ for $N=3$ up to $8$ as a function of the angular momentum $l-l_0$. For
all $N$ one can clearly identify a branch of chiral excitations propagating to the right. The analytical predictions are indicated by the dimensions of the
$SU(N)$ irreducible representations. For all $N$ the numerical data for the first three sectors ($l-l_0=0,1,2$) is in full agreement with the analytical 
predictions. The splitting between the multiplets at a given value of $l$ is expected to vanish as $N_s$ grows, and the spectrum should become linear 
with a certain edge state velocity. In summary the analysis of the structure of the edge excitations performed here confirms the $SU(N)_1$ WZNW CFT predictions
and thus strengthens the case for abelian $SU(N)$ chiral spin liquids in the model Hamiltonian~\eqref{eqn:Heisenberg}.

\paragraph{Experimental considerations --}
With the recent realization of the Mott-crossover regime in 3D optical lattices with fermionic Ytterbium atoms~\cite{Taie2012,Hofrichter2015}
the prospect for the realization of strongly correlated $SU(N)$ quantum magnetism is becoming bright. Our proposal for triangular lattices
builds on ingredients which have been demonstrated separately: the possibility to realize Mott insulators in optical lattices, and to create static artificial gauge 
fields in an optical lattice (for alkaline atoms)~\cite{PhysRevLett.111.185301,PhysRevLett.111.185302}. Beside, working with the triangular lattice is a big advantage because the 3-site permutation term is the first and only term to appear to third order perturbation theory starting from the Hubbard model with one particle per site, by contrast to e.g. the square and honeycomb
lattice, where they appear at order 4 and 6 respectively, and are not the first corrections. The chiral phase typically appears for 
$\theta \simeq 0.3$, which, using the perturbation expressions of $J=2t^2/U$ and $K_3=6t^3/U^2$, corresponds to $t/U \simeq 0.1$. This 
might be small enough to be still in the Mott insulating phase, and to ensure that higher order corrections are negligible. In future studies one
might also relax the $\pi/2$ flux per plaquette condition, and explore the extent of the expected stability region of the $SU(N)$ CSL phases.

Several interesting questions need to be addressed in future work. For example, is it possible to directly engineer the required three
site exchange terms in Hamiltonian~\eqref{eqn:Heisenberg} using sophisticated quantum optics schemes? There is hope that the
current activity on lattice gauge-theory implementations will bring techniques to address this question. Another intriguing question regards
the detection of $SU(N)$ chiral spin liquid edge states in actual experiments, for example using spectroscopic techniques for small droplets,
or braiding protocols for the abelian anyons~\cite{HermeleGurarie2011}.

\begin{acknowledgements}  
The authors acknowledge P.~Corboz, M.~Hermele, T.~Quella, A.~Sterdyniak, H.-H.~Tu and Hongyu Yang for useful discussions. This work
has been supported by the Swiss National Science Foundation, the JSPS KAKENHI Grant Number 2503802, the Hungarian OTKA Grant No. K106047
and by the Austrian Science Fund FWF (F-4018-N23 and I-1310-N27/DFG-FOR1807).
\end{acknowledgements}

\bibliographystyle{apsrev4-1}
\bibliography{biblio,chiralspinliquid,SU6_refs}
\begin{table*}[b]
\begin{tabular}{|c|c|c|c|c|c|}
\hline
N in SU(N)  &$N_s \mod{N}$ & $l=0$ & $l=1$ & $l=2$ \\
\hline
\hline
$2$ & $1$ & ${\tiny \yng(1)}\ (2)$ & ${\tiny \yng(1)}$  & ${\tiny \yng(1)} \oplus {\tiny \yng(3)\ (4)}$ \\
\hline
$3$ & $1$ &${\tiny \yng(1)}\ (3)$ & ${\tiny \yng(1) \oplus \tiny \yng(2,2)}$ ($\bar6$) & {$\tiny 2\times \yng(1)\oplus  \yng(2,2) \oplus \tiny \yng(3,1)$ ($15$)}  \\
\hline
$4$ & $3$ &${\tiny \yng(1,1,1)}\ (\bar{4})$ & ${\tiny \yng(1,1,1) \oplus \tiny \yng(2,1)}$ ($20$) &{$\tiny 2\times \yng(1,1,1)\oplus  2\times \yng(2,1) \oplus \tiny \yng(3,2,2)$ ($\bar{36}$)}  \\
\hline
$5$ & $4$ &${\tiny \yng(1,1,1,1)}\ (\bar{5}) $& ${\tiny \yng(1,1,1,1)}  \oplus{\tiny \yng(2,1,1)}\ (45)$ &{$\tiny 2\times \yng(1,1,1,1)\oplus  2\times\yng(2,1,1) \oplus \yng(2,2)\ (50)\oplus \yng(3,2,2,2)\ (\bar{70})$ }   \\
\hline
$6$ & $1$ &${\tiny \yng(1)}\ (6)$ & ${\tiny \yng(1) \oplus \tiny \yng(2,2,1,1,1)}\ (\bar{84})$& ${\tiny 2\times \yng(1) \oplus 2\times \yng(2,2,1,1,1)  \oplus \yng(3,1,1,1,1)\ (120) \oplus \yng(2,2,2,1)\ (\bar{210})}$ \\
\hline
$7$ & $5$ &${\tiny \yng(1,1,1,1,1)}\ (\bar{21})$& ${\tiny \yng(1,1,1,1,1) \oplus \tiny \yng(2,2,2,2,2,2)}\ (\bar{28}) \oplus\tiny \yng(2,1,1,1)\ (224)$ & 
${\tiny 3\times \yng(1,1,1,1,1) \oplus \yng(2,2,2,2,2,2) \oplus 2\times \yng(2,1,1,1) \oplus \yng(2,2,1)\ (490) \oplus \yng(3,2,2,2,2,1)\ (\bar{735})}$\\
\hline
$8$ & $3$ &${\tiny \yng(1,1,1)\ (56)}$ & ${\tiny \yng(1,1,1) \oplus \tiny \yng(2,1)\ (168) \oplus \yng(2,2,2,2,1,1,1)\ (\bar{504})}$&
${\tiny 3\times \yng(1,1,1) \oplus 2\times \yng(2,1) \oplus 2\times \yng(2,2,2,2,1,1,1) \oplus \yng(2,2,2,2,2,1)\ (\bar{1008}) \oplus \yng(3,2,2,1,1,1,1)\ (2800)}$\\
\hline
$9$ & $1$ & ${\tiny \yng(1)\ (9)}$ & ${\tiny \yng(1) \oplus \yng(2,2,1,1,1,1,1,1)\ (\bar{315})}$ &
${\tiny 2\times \yng(1) \oplus 2\times \yng(2,2,1,1,1,1,1,1) \oplus \yng(3,1,1,1,1,1,1,1)\ (396) \oplus \yng(2,2,2,1,1,1,1)\ (\bar{2700})}$
\\
\hline
\end{tabular}
\caption{
The three first angular momentum sectors of the chiral edge mode of the $N_s=19$ droplet for SU(N)$_1$. The
number $(N_s\mod{N})$ selects the anyonic sector (primary field of the CFT) which consequently determines the  edge spectrum.
}
\label{tab:counting_sun_1_Ns_19}
\end{table*}
\newpage
\section*{Supplementary material}
\subsection*{$SU(N)_1$ WZWN predictions for the chiral edge states}

In Tab.~\ref{tab:counting_sun_1_Ns_19} we explicitly list the expected $SU(N)$ irreducible representations with
their multiplicity for the first three excitation levels $l=0$ (primary field) and $l=1,2$ (first two descendant levels) of a chiral 
$SU(N)_1$ WZNW conformal field theory. The primary field for each $N$ is dictated by the open boundary clusters size $N_s=19$ 
via the length $(N_s \mod N)$ of the single-column young diagram at $l=0$. We have derived these results using
a successive $SU(N)$ coupling sequence with the adjoint representation starting from the irreducible representation at $l=0$ 
and subsequent the null-vector elimination based on the $SU(N)$ counting rule restrictions listed in Tab.~\ref{tab:counting}.
This simplified procedure uses the fact that the $SU(N)_1$ CFT can also be seen as particular $SU(N)$ invariant combination 
of $N-1$ Luttinger liquid CFTs (thus the $SU(N)_1$ WZNW CFT central charge $c=N-1$).

\begin{table}[b]
\begin{tabular}{|c|c|c|}
\hline
SU(N) &  counting rule & OEIS identifier\\
\hline
\hline
SU(2) & $1, 1, 2, 3, \ldots$  &\href{https://oeis.org/A000041}{A000041}\\
\hline
SU(3) &  $1, 2, 5, 10, \ldots$ &\href{https://oeis.org/A000712}{A000712}\\
\hline
SU(4) & $1, 3, 9, 22, \ldots$ &\href{https://oeis.org/A000716}{A000716}\\
\hline
SU(5) & $1, 4, 14, 40, \ldots$ &\href{https://oeis.org/A023003}{A023003}\\
\hline
SU(6) & $1, 5, 20, 65, \ldots$ &\href{https://oeis.org/A023004}{A023004}\\
\hline
SU(7) & $1, 6, 27,98, \ldots$ &\href{https://oeis.org/A023005}{A023005}\\
\hline
SU(8) & $1, 7, 35, 140,\ldots$ &\href{https://oeis.org/A023006}{A023006}\\
\hline
SU(9) & $1, 8, 44, 192,\ldots$ &\href{https://oeis.org/A023007}{A023007}\\
\hline
SU(10)&$1, 9, 54, 255,\ldots$&\href{https://oeis.org/A023008}{A023008}\\
\hline
\end{tabular}
\caption{$SU(N)$ counting rules used to derive the results in Tab.~\ref{tab:counting_sun_1_Ns_19}.}
\label{tab:counting}
\end{table}

\end{document}